\documentclass[]{spiejour}  
\usepackage{amsmath}  
\usepackage[]{graphicx}
\usepackage{epstopdf,mathrsfs,amsmath,amsfonts,textcomp}
\DeclareMathOperator{\erf}{erf}

\title{A compact analytic solution describing optoacoustic phenomenon in absorbing fluid}

\author{Luisiana Cundin\supit{a}, William Elliott\supit{b}, Norman Barsalou\supit{c} \& Shannon Voss\supit{d}}

\authorinfo{\supit{a}Conceptual Mindworks, Inc.\ \supit{b}Henry M. Jackson Foundation\ \supit{c}NAMRU-SA Directed Energy Biomedical Research (DEBR) Department\ \supit{d}Head, NAMRU-SA Directed Energy Biomedical Research (DEBR) Department}

\begin{document} 
  \maketitle 

\begin{abstract}
Derivation of an analytic, closed-form solution for Q-switched laser induced optoacoustic phenomenon in absorbing fluid media is presented. The solution assumes spherical symmetry as well for the forcing function, which represents heat deposition from Q-switched lasers. The Green's solution provided is a suitable kernel to generate more complex solutions arising in optoacoustics, optoacoustic spectroscopy, photoacoustic and photothermal problems.  
\end{abstract}


\section{Introduction}
Optoacoustic phenomenon is the formation of sound waves in absorbing media by means of either a modulated or pulsed light source \cite{Patel, Rosencwaig, MJAdams}. Regardless of the impulse, an acoustic wave involves the physical vibration of the constituent parts comprising the material and the cumulative propagation of the wave enjoins all material properties in total; as a consequence, the information contained within the wave envelopes all the fundamental physical properties of the material. The difficulty rest in deciphering signals detected by various means, where some detection methods measure the physical vibration and other methods exploit derivative thermodynamic relations. 

Photoacoustic and photothermal spectroscopy are two separate detection methods for the same physical phenomenon, that is, both are semantic nomenclatures for optoacoustic phenomenon. The generation of an acoustic pressure wave in ponderable media can be detected either by measuring the amplitude of the stress vector (the physical displacement) or indirectly by measuring the degree by which some intrinsic physical parameter changes. Direct measurement of the stress vector can be monitored by utilizing an electric transducer and this detection method constitutes photoacoustic phenomenon. Although many intrinsic characteristics change during the disturbance, measuring the degree by which the refractive index changes constitutes the photothermal detection method. This method makes use of the refractive index change caused by the pressure wave to deflect a probe laser beam. The probe beam is projected onto a differential photodetector, which differences the magnitude of several segmented regions of the 
photodetector. The probe beam deflects as the acoustic wave passes at point in space; furthermore, both the direction and magnitude of the deflection angle is proportional to the shape of the refractive index change caused by the passing pressure wave.

Laser technology enables ultra-short pulses of laser light, of which Q-switched lasers are an example. The physical interaction of photon energy and ponderable media is quite complex. The particular mode of interaction is dependent upon the wavelength of light, pulse energy, beam characteristics, pulse duration and material constants, such as thermal conductivity, density, \&c. Despite the many complexities of interaction, Lambert's absorption coefficient is an empirical measure of the coupling between light of a given frequency and media of interest. For high frequency radiation, the general mode of interaction is to excite both atomic and molecular vibrational and/or electronic transitions. Excess energy absorbed by atoms and molecules are expelled through molecular relaxation processes. 

The optoacoustic phenomenon is an isentropic process, where the condensation (compression) of the media at the leading edge of the wavefront causes the material to rapidly increase in temperature; similarly, as the pressure wave passes, the subsequent rarefaction (expansion) allows for the cooling of the material. This would be a reversible adiabatic process if it were not for nonconservative forces dissipating the energy as the acoustic wave propagates; albeit, the convolution of such nonconservative forces temporally and spatially allow atomic and/or molecular characteristics to be imprinted onto the resultant acoustic wave, which are then detected, identified and deciphered. Deconvolution of the resulting signal is confounded by the propagation of the disturbance through the volume of interest. Subsequent reflections from the container walls, material interaction and nonconservative forces from chemical constituents increases the complexity of the resultant wavefront; thus, detection of the initial 
disturbance is 
most efficacious.

Optoacoustics and optoacoustic spectroscopy is undoubtedly a very rich and complex technique and cannot be discussed comprehensively in such few words; but, the purpose of the present expos\`{e} is to derive a compact, closed-form analytic solution for the coupled partial differential system describing this complex phenomenon. By compact is meant an analytic expression relating all requisite parameters involved in generating an acoustic wave from a pulsed laser source in as few terms as possible. Thus, the intent is to present a functional that can serve as a kernel solution to generate more complex solutions, to be used in circumstances involving greater specificity in geometry, material properties, \&c. Discussions will concentrate on the generation and behavior of the resulting acoustic wave; the topic of detection is deferred.  

\section{Assembly}
For incompressible irrotational fluid flow, pressure $p$ is proportional to the time rate of change of a scalar potential $\Phi$ which describes the physical displacement of particles and fluid, encapsulated by equation (1) \cite{Fetter,LandauFD}. The scalar potential is related to the time rate of change of the temperature T in the fluid through the wave equation, where the characteristic speed of a resulting acoustic disturbance through the medium is represented by $v$, enjoined by equation (2). The origin of any temperature disturbance results from a forcing function ($\mathrm{S}$) acting on the fluid, defined by equation (3). Tracing the relationship the scalar potential has with pressure, it can be seen that pressure is proportional to the second-order time derivative of temperature within the fluid, hence, pressure is directly related to the particle and/or fluid acceleration. Temperature is proportional to the absorbed portion of the laser's emitted time averaged radiant energy per volume; 
equivalently, the temperature is proportional to the absorbed portion of the emitted spectral irradiance [Watts per volume]. The temperature in the fluid is represented by the standard forced heat equation, i.e. equation (3). The coupled partial differential system of equations, equations (1-3), model pressure waves initiated by thermal deposition from a forcing function acting on the fluid. 
\begin{align}\label{pressure}
 p=&-\rho_0\frac{\mathrm{d}\Phi}{\mathrm{d}t}\\\label{velocity_potential}
\left(\nabla^2-\frac{1}{v^2}\frac{\mathrm{d}^2}{\mathrm{d}t^2}\right)\Phi=&-\beta\frac{\mathrm{d}\mathrm{T}}{\mathrm{d}t}\\\label{heat}
\rho_0 C_p\frac{\mathrm{d}\mathrm{T}}{\mathrm{d}t}=&\,\kappa\nabla^2 \mathrm{T}+\mathrm{S}
\end{align}

The forced heat equation describes the time rate of change in temperature, scaled by the fluid density ($\rho_0$) and the heat capacity ($C_p$), to be equal to the sum of the spatial curvature times the thermal conductivity $\kappa$ of the medium and the forcing function ($\mathrm{S}$). The thermal conductivity is assumed isotropic within the media, thus, the heat equation model is appropriate for absorbing fluid. Heat deposition from an external laser source is the first physical phenomenon to be tackled in the treble partial differential system, equations (1-3). 

The force $\mathrm{S}$ causing the initial thermal disturbance could be attributed to many sources, but we concentrate upon disturbances caused by Q-switched laser pulses. Inspection of the forced heat equation shows there are two contributions for rapid, concentrated heat deposition; specifically, either the spatial curvature of the deposition is large, meaning the spatial gradient is large, or the time derivative of heat deposition is large. In general, the amplitude of the resulting pressure wave is directly proportional to the power density of the impinging laser pulse. 

The impinging light is focused on the absorbing media and the absorbed energy raises the temperature of the media. Photons entering the media are absorbed according to Beer-Lambert-Bouguer's law of absorption, where absorption of energy can be treated omnidirectional; thus, the forcing function may be treated as spherically symmetric. The original spatial domain will be mapped to the Fourier codomain by employing three-dimensional Fourier transforms (spherical transforms), where the spherical radial variable ($r$) is mapped to the codomain variable ($s$), \emph{ergo} $r\mapsto s$. Transforms will be represented by either Bracewell's symbolism for a Fourier transformation ($\subset$) or by the following symbol: $\mathscr{F}_{\mathbb{R}^3}$ \cite{Bracewell}. If the immediate assumption of spherical symmetry be perceived a limitation, it is possible to break the spatial symmetry at a latter point in time by employing the convolution theorem. 

The forcing function $\mathrm{S}$ describes the manner in which an impinging laser would interact with some ponderable media. It is common to consider the forcing function separable in space and time; this greatly facilitates prehending a solution for the partial differential equation. Additionally, stochastic independence is often assumed for an incident laser beam, given the statistical nature of photons. The source function $\mathrm{S}$ is separated into the product of two functions, where $f(t)$ defines the time dependent behavior of the source, $\mathrm{P}(r)$ describes the spatial distribution of the laser beam.
\begin{equation}\label{source_assumption}
 \mathrm{S}(r,t)\Rightarrow \mathrm{P}_0f(t)\mathrm{P}(r)\subset \mathrm{P}_0f(t)\mathrm{P}(s);\ r\mapsto s
\end{equation}

The above equation shows the decomposition of the source function into the spatial and temporal components of the incident laser pulse. The three-dimensional Fourier transform of the spatial distribution of the incident laser intensity $\mathrm{P}(s)$ is multiplied by a suitable constant $\mathrm{P}_0$ to scale the amplitude of the incident laser energy. 

Let the spatial distribution for the light source remain arbitrary for the moment and concentrate upon the particular solution for the forced heat equation. The Fourier transform of the forced heat equation, in combination with the source decomposition specified, provides a time dependent ordinary differential equation, which can be readily solved \cite{Boyce}. The complementary solution, commonly referred to as Green's heat kernel, describes the decay of an initial heat distribution and is not shown in the solution, because the initial heat profile is assumed regular and quiescent. The particular solution describes the contribution of heat attributable to the external source acting upon the media over some duration of time, \textit{viz}.:
\begin{equation}\label{particular}
 \mathrm{T}(s,t)=\frac{\alpha\mathrm{P}_0\mathrm{P}(s)}{\kappa}e^{-\alpha(2\pi s)^2t}\int{e^{\alpha(2\pi s)^2t^\prime}f(t^\prime)\,\mathrm{d}t^\prime}
\end{equation}

The particular solution of the forced heat equation in the Fourier codomain involves the superpositioning of both fundamental solutions to the ordinary differential equation. Integrating Green's unbounded heat kernel multiplied by the specific time dependence for the external source of light describes the accumulation of heat within the media. The integral is then multiplied by Green's heat kernel, the diffusivity $\alpha$, the magnitude of the laser intensity $\mathrm{P}_0$ and the Fourier transformed spatial distribution of the laser light $\mathrm{P}(s)$, and finally, divided by the thermal conductivity $\kappa$. Diffusivity describes the rate at which heat will flow in a media and is defined as the division of the thermal conductivity by both the material density and heat capacity. The specific spatial distribution of the laser intensity $\mathrm{P}(s)$ will be ignored presently, because this term is time independent and therefore may be treated as a constant. 

Regarding the time function $f(t^\prime)$, this describes how the light source is modulated in time, but the specific modulation is immaterial. Duhamel's \emph{principle} can be applied later and convolve the solution with any specified modulation; as a consequence, let the time function $f(t^\prime)$ be constant over the interval $[0,t]$, which will set the limits on the time integration. Completing the time integration leads to a compact solution, \textit{viz}.:
\begin{equation}\label{fourtemp}
 \mathrm{T}(s,t)=\frac{\mathrm{P}_0\mathrm{P}(s)}{\kappa}\left(\dfrac{1}{(2\pi s)^2}-\dfrac{e^{-\alpha(2\pi s)^2t}}{(2\pi s)^2}\right)
\end{equation}

The temperature solution is rather general in nature, the specific spatial distribution of the source is multiplied over the kernel solution in Fourier space. The second term (within brackets) represents the upper limit of the time integral and is left open-ended for arbitrary time $t$. In the limit of infinite time, the second term vanishes, thus, showing the purpose this term has for regulating the time deposited heat within the material. Division by the thermal conductivity $\kappa$ scales the magnitude of the resulting temperature. 

Considering the spatial distribution function $\mathrm{P}(s)$ a constant, then the solution presented in equation (\ref{fourtemp}) is essentially a Green's function for the corresponding forced ordinary differential equation. To see the influence an incident laser possessing beam characteristics and material interaction has on heat deposition, then both the beam width ($\omega_0$) and the penetration depth ($\gamma$) must be brought into the solution. The penetration depth is inversely proportional to Lambert's absorption for a material and is measured in units of length; equivalently, the penetration depth is inversely proportional to the absorption coefficient, i.e. $a=1/\gamma$. The beam width represents the axial distance for the Full Width Half Maximum (FWHM) of a Gaussian beam. Appropriate division by a rational polynomial of the incident Gaussian beam is taken to ensure the Fourier transform has unit area, \textit{viz}.:
\begin{equation}\label{spatial}
 \mathrm{P}(s)\equiv\mathscr{F}_{\mathbb{R}^3}\Big\{\mathrm{P}(r)\Big\}\equiv\mathscr{F}_{\mathbb{R}^3}\Bigg\{\frac{\exp\left(\frac{-r^2}{\gamma^2+\omega_0^2}\right)}{(\gamma^2+\omega_0^2)^{3/4}}\Bigg\}= \frac{e^{-\pi^2(\omega_0^2+\gamma^2) s^2}}{\pi^{3/2}}
\end{equation}

It can be seen that the incident light has a constant power profile, it is most efficient to control the intensity by an arbitrary constant rather than parameters used to describe beam characteristics. Using equation (\ref{spatial}) to describe the impinging laser light, a specific solution for the temperature can be derived, which includes the beam characteristics, \textit{viz}.:
\begin{equation}\label{xspacefouriersolution}
 \mathrm{T}(r,t)=\frac{\mathrm{P}_0}{\sqrt{\pi}\pi^{2}\kappa}\frac{1}{r}\left\{\erf\hspace{-2pt}\left(\frac{\sqrt{\pi}r}{\sqrt{\gamma^2+\omega_0^2}}\right)-\erf\hspace{-2pt}\left(\frac{\sqrt{\pi}r}{\sqrt{4\alpha t+\gamma^2+\omega_0^2}}\right)\right\}
\end{equation}

The overall temperature is controlled by the power of the incident light source; $\mathrm{P}_0$, measured in units of power (Watts). The temperature in the media is mitigated by several factors, namely, the geometric dependence ($1/r$), the beam width ($\omega_0$) and the penetration depth ($\gamma$). The algebraic sum of the two error functions represents the cumulative energy absorbed by the system and that power is scaled by the width of the incident laser beam, the absorption of energy governed by the penetration depth, and the period of time the laser is active. The temperature reaches it's maximum as time progresses; this is the particular solution for deposited energy from an impinging light source. If diffusion should also be modeled, then the accumulated heat, equation (\ref{xspacefouriersolution}), must be the leading coefficient for Green's heat kernel, which describes the dissipation of energy within the system. Due to the \emph{maximum principle}, one cannot represent the rise and dissipation 
of heat on the same timeline; rather, one is forced to segment the two time regimes \cite{Weinberger}. For energy deposition over large time periods, diffusion becomes the predominant process; but, for short time regimes, acoustic wave generation predominates.  

The spatial distribution of the light source determines the spatial relationship temperature has with the media; as the width of the incident beam is increased, more of the incident energy is distributed over a wider volume and this effectively reduces the magnitude of the temperature reached over any period of time. Similarly, the magnitude of the temperature is inverse to the penetration depth; hence, the less absorptive the media is to a particular frequency of light, the lower the resulting temperature within the media. The contribution of either the beam width or the penetration depth can be ignored by simply considering an infinitely absorbing media ($\gamma\ll 1$) and an infinitely tight beam ($\omega_0\ll 0$); then the deposition of heat is governed solely by the product of the diffusivity and the pulse width. For sufficiently short enough pulses, measured relative to diffusivity, the ability to realize \textit{thermal confinement} becomes possible. Thermal confinement refers to the process of 
dumping significant enough energy locally within the media, rapidly enough as to cause nonlinear effects. 

Now that the temperature is known, we place attention on the scalar potential. The scalar potential function, equation (2), is essentially the forced wave equation and yields a second order differential equation after transforming the spatial coordinates to the Fourier codomain; once again, assuming spherical symmetry. The forcing function for the scalar potential is equal to the time derivative of the temperature function scaled by the thermal expansion coefficient $\beta$. The derivative of the forcing function does not involve the spatial variables regardless of what domain is considered, either the original domain or the Fourier codomain; thus, the time derivative of equation (\ref{fourtemp}) yields equation (\ref{forcingvel}).
\begin{equation}\label{forcingvel}
 F(t)\equiv -\beta\frac{\mathrm{d}}{\mathrm{d}t}\mathrm{T}(s,t)=-\beta\frac{\alpha\mathrm{P}_0\mathrm{P}(s)}{\kappa}e^{-\alpha(2\pi s)^2t}
\end{equation}

The solution for the scalar potential is a sum of two complementary solutions and the particular solution; we are only interested in the particular solution. The complementary solution provides the behavior of the space itself, also, any constant waves existent within the media prior to the action of the forcing function are not of any interest. The particular solution $Y(t)$ involves both complementary solutions, $y_1(t)$ and $y_2(t)$, the forcing function $F(t)$ and the Wronskian $W(t)$, \textit{viz}.:
\begin{equation}
 Y(t)=\,y_1(t)\int{\frac{y_2(t)F(t)}{\mathrm{W}(t)}\,\mathrm{d}t}+y_2(t)\int{\frac{y_1(t)F(t)}{\mathrm{W}(t)}\,\mathrm{d}t}
\end{equation}

With complementary solutions:
\begin{equation}
 y_1(t)=e^{-2\pi isvt},y_2(t)=e^{2\pi isvt}
\end{equation}

The Wronskian W$(t)$ describes linear independence for the solution and, by inspection of equation (\ref{wronskian}), the solution is independent for all time except for the origin of the transform variable or a zero velocity wave, \textit{i.e.} $\{s,v\}=0$. An acoustic wave with a null velocity must be rejected. An association with the origin is implicit, because the initial condition is considered quiescent; hence, there can be no acoustic wave if the impulse causes a constant disturbance throughout space. Because the field is considered constant prior to the action of the forcing function, that is, the incident laser light, we must disregard the solution dealing with time zero.
\begin{equation}\label{wronskian}
 \mathrm{W}(t)=\mathrm{det}\left|\begin{array}{cc}
                 y_1(t) & y_2(t)\\
y_1^\prime(t) & y_2^\prime(t)\\
                \end{array}\right|=4\pi ivs=0
\end{equation}

The particular solution is generated by considering both the incoming and outgoing forms of the acoustic wave. Since the outgoing wave, emanating from a source or disturbance is sought, the particular solution $Y(t)$ takes on the form as shown in equation (\ref{partsoln}). Integrating the forcing function with the incoming wave form describes how the disturbance accumulates over time. The time integral is multiplied by the outgoing wave form, describing the transmission of any accumulated energy out from the source. 

Now the limits of integration are over the interval of time the light source should cause heat deposition within the media, which has already been stated to range from zero to arbitrary time \textit{t}, i.e. $[0,t]$. Because the quiescent period prior to external perturbation by the incident light is inconsequential, as evidenced by the Wronskian, then all that is required is an end point in time when the laser pulse is shut off. The restraint for the limits of integration can be succinctly represented by implanting a Dirac delta function shifted by the pulse width ($t_0$). This is an approximation applied to the time integration for the formation of the disturbance and will not accurately describe the formation of the acoustic wave itself. The approximation will become better as the pulse width in time is shortened. Nevertheless, the chosen method of describing the time dependence is applicable no matter the time period for the pulse as long as the range for the time variable is truncated accordingly.
\begin{equation}\label{partsoln}
 Y(t)=\frac{e^{-2\pi isvt}}{4\pi ivs}\int_{-\infty}^{\infty}{e^{2\pi isvt^\prime}e^{-\alpha(2\pi s)^2 t^\prime}\delta(t^\prime-t_0)\,\mathrm{d}t^\prime}
\end{equation}

The leading coefficients are not shown in the particular solution $Y(t)$ for brevity, such as $\mathrm{P}_0$, $\alpha$, \&c. Taking the limit of infinite time, Green's heat kernel diminishes and the thermal contribution to the resulting wave approaches zero, canceling any acoustic wave; thus, the impulse must be chopped or modulated in time to cause an appreciable acoustic wave. This justifies the approximation for the integral by truncating to the upper time limit for the laser pulse width. In the limit of the pulse width approaching zero, the integral approaches unity, thus, the resulting acoustic wave is that from the most fundamental impulse possible, i.e. a Dirac delta function. Shifting the placement of this impulse in time approximates the accumulation of deposited heat over pulse width. After completing the integration, the solution is a product of a sinusoidal wave and an exponential, also, division by the Wronskian, \textit{viz}.:
\begin{equation}\label{mainsoln}
 Y(t)=\frac{ie^{-2\pi isv(t-t_0)}}{4\pi vs}e^{-\alpha(2\pi s)^2 t_0}
\end{equation}

The denominator indicates dispersion of the wave, where the magnitude of the resulting wave is scaled by the spatial frequency component, thus, the amplitude of the high frequency wave component is more swiftly reduced relative to the slower longer period waves. 

It is now a matter of returning the particular solution for the scalar wave equation to the original spatial domain; the inverse Fourier transform involves complex integration techniques and both real $\Re$ and imaginary $\Im$ parts are shown in equations (\ref{sinesoln}) and (\ref{cosinesoln}).
\begin{align}\nonumber
 \mathscr{F}_{\mathbb{R}^3}^{-1}\Bigg\{
\frac{\sin(2\pi sv(t-t_0))}{4\pi sv}&e^{-\alpha(2\pi s)^2t_0}\Bigg\}=\\\label{sinesoln}
&\frac{1}{4\pi^{3/2} vr\sqrt{4\alpha t_0}}\Bigg\{
e^{\frac{-(r-v(t-t_0))^2}{4\alpha t_0}}-
e^{\frac{-(r+v(t-t_0))^2}{4\alpha t_0}}\Bigg\}
\end{align}

The inverse Fourier transform of the imaginary part ($\Im$) of the particular solution $\sim$
\begin{align}\nonumber
\mathscr{F}_{\mathbb{R}^3}^{-1}\Bigg\{
 \frac{\cos(2\pi sv(t-t_0))}{4\pi sv}e^{-\alpha(2\pi s)^2t_0}\Bigg\}=&\\\nonumber
\frac{1}{4\pi^{3/2} vr\sqrt{4\alpha t_0}}\Bigg\{
e^{\frac{-(r+v(t-t_0))^2}{4\alpha t_0}}&\mathrm{\textit{erf\hspace{0.3pt}i}}\hspace{-2pt}\left[\frac{r+v(t-t_0)}{\sqrt{4\alpha t_0}}\right]+\\\label{cosinesoln}
&e^{\frac{-(r-v(t-t_0))^2}{4\alpha t_0}}\mathrm{\textit{erf\hspace{0.3pt}i}}\hspace{-2pt}\left[\frac{r-v(t-t_0)}{\sqrt{4\alpha t_0}}\right]\Bigg\}
\end{align}

Where the imaginary error function is defined as such $\sim$
\begin{equation}
 \mathrm{\textit{erf\hspace{0.3pt}i}}(z)=-i\erf(iz)
\end{equation} 

Both solutions are implicitly valid for the entire real line for both spatial and temporal coordinates; but, due to the principle of \emph{causality}, the appropriate range for the spatial coordinates should not exceed the distance allowed by the speed of the acoustic wave multiplied by the time traveled, \textit{viz}.:
\begin{equation}
 \left\{r|r\leq v(t-t_0),r\in [0,\infty]\right\},\, \left\{t|t\in [t_0,\infty]\right\}
\end{equation} 

There are two solutions generated from equation (\ref{mainsoln}), each represents a 90\textdegree\ phase difference. The sinusoidal solution implies a disturbance arising out of phase with the perturbing force causing the disturbance; contrary, the cosinusoidal solution describes a disturbance concurrent with the perturbation. It is the cosinusoidal solution we are most interested in. Both solutions show the amplitude of the resulting acoustic wave is inversely proportional to the speed of the wave in the media ($v$) and the geometric dispersion ($1/r$). The product between diffusivity and pulse width ($\alpha t_0$) determines the degree of \textit{thermal confinement}. The larger the diffusivity, the swifter the thermal conductivity of the media, the smaller the amplitude for the resulting acoustic wave. Diffusivity restrains the pulse width of the impinging light to enforce thermal confinement, where for short enough pulses ablation can occur. In contrast, for pulse widths many orders of 
magnitude larger than the diffusivity, the material allows deposited heat to be diffused away, thus, quenching any anticipated non-linear effects.

Considering the specific parameters attributed to the light source, equation (\ref{spatial}), it is a simple matter to fold this description into the known inverse Fourier transform solution represented by the cosinusoidal part of the wave solution, equation (\ref{cosinesoln}); moreover, this combined solution will be signified by the symbol $\mathrm{K}(r,t)$ for later reference.
\begin{align}\nonumber
 \mathrm{K}(r,t)\equiv\mathscr{F}_{\mathbb{R}^3}^{-1}\Bigg\{\frac{e^{-\pi^2 (\gamma^2+\omega_0^2) s^2}}{\pi^{3/2}}&\frac{\cos(2\pi sv(t-t_0))}{4\pi sv}e^{-\alpha(2\pi s)^2t_0}\Bigg\}=\\\nonumber
\frac{1}{4\pi^3 vr\sqrt{4\alpha t_0+\gamma^2+\omega_0^2}}&\Bigg\{
e^{\frac{-(r+v(t-t_0))^2}{4\alpha t_0+\gamma^2+\omega_0^2}}\mathrm{\textit{erf\hspace{0.3pt}i}}\hspace{-2pt}\left[\frac{r+v(t-t_0)}{\sqrt{4\alpha t_0+\gamma^2+\omega_0^2}}\right]+\\\label{soln}
e^{\frac{-(r-v(t-t_0))^2}{4\alpha t_0+\gamma^2+\omega_0^2}}&\mathrm{\textit{erf\hspace{0.3pt}i}}\hspace{-2pt}\left[\frac{r-v(t-t_0)}{\sqrt{4\alpha t_0+\gamma^2+\omega_0^2}}\right]\Bigg\}
\end{align}

Analyzing the solution, the amplitude of the resulting acoustic wave is inversely proportional to the distance (\textit{r}), the product ($\alpha t_0$), the penetration depth ($\gamma$), and finally, the radial width of the laser pulse ($\omega_0$). In general, for either a weakly absorbing material, $\gamma\gg 1$, or for greatly dispersed beams, $\omega_0\gg 1$, the resulting amplitude of the acoustic wave is vanishingly small; thus, both of these parameters control the ability to cause an acoustic wave of any significant amplitude. But, taking the limit of both of these parameters to the other extreme, infinitely absorbing and infinitely confined beam ($\{\gamma,\omega_0\}\rightarrow 0$), the ability to cause thermal confinement is the limiting factor responsible for the amplitude of the acoustic wave.  

The real ($\Re$) part of the functional, $\mathrm{K}(r,t)$, is what is of interest. Keeping in mind the constraints \emph{causality} placed upon the range for all relevant parameters, the functional provides a compact, closed-form analytic solution for the scalar potential wave arising from Q-switched laser pulses. The solution for the scalar potential ($\Phi$) can now be constructed, where earlier ignored leading coefficients can be reintroduced, \textit{viz}.: 
\begin{equation}
 \Phi=\frac{\alpha\beta v^2\mathrm{P}_0}{\kappa}\mathrm{K}(r,t)
\end{equation}

Based upon the known functional for the scalar scalar potential: the velocity $v(t)$, pressure $p(t)$ and density $\rho(t)$ of the resulting acoustic wave are all proportional to either the spatial gradient ($\nabla$) or the time derivative ($d/dt$) of the scalar potential, \textit{viz}.: 
\begin{align}
\mathrm{velocity:}\ v(t) &=-\nabla\Phi\\
\mathrm{pressure:}\ p(t) &=-\rho_0\frac{\mathrm{d}\Phi}{\mathrm{d} t} \\
\mathrm{density:}\ \rho(t) &=-\frac{\rho_0}{v^2}\frac{\mathrm{d}\Phi}{\mathrm{d} t}
\end{align}

\section{Closing}
A need for a compact, closed-form analytic solution describing laser induced optoacoustic wave generation was identified. The most fundamental and efficient solution was sought to serve as a kernel for further complications in modeling \& simulation and/or theoretical investigations. The system of partial differential equations used to describe the absorption of laser energy, subsequent heat generation, and finally, formation and propagation of an acoustic pulse requires multiple integrations; furthermore, it is most efficient to incorporate most of the information contained in the treble differential system in as succinct a formulation as possible. A suitable solution for the scalar potential enables calculating the velocity, pressure and density of the resulting acoustic wave. A kernel solution, as the nomenclature implies, serves as a seed or kernel for further complication, where specific temporal or spatial constraints can be entertained.  

The analytic solution presented assumes both spherical symmetry in physical space and rectangular time dependence. Duhamel's \emph{principle} allows further refinement of the time dependence by appropriate convolution of the kernel solution with some suitable time function; thereby, enabling temporal modulation of the impinging laser pulse. For the kernel solution provided, the beam characteristics attributed to the impinging laser pulse was assumed to be spherically symmetric; but, if further complication be desired, convolving the kernel over some specific spatial distribution can be performed, where appropriate consideration of the spatial basis set is taken. In either case, application of Duhamel's \emph{principle} or spatial convolution, both of these operators should be applied directly to the scalar potential ($\Phi$) solution and applied before any other derivative is applied, specifically, before applying those derivatives indicated in equations describing various physical observables, i.e. velocity,
 pressure. 

The mission of NAMRU-SA/Directed Energy Biomedical Research Department is to investigate nonionizing radiation bioeffects and understand the potential risks to human health and safety. Q-switched laser pulses can cause damage to the retina, including hemorraghing caused by a laser induced optoacoustic pulse. Pulsed laser energy is unique in that the radiant energy is very low, but the radiant power of the pulse is very large; thus, the temporal nature of short pulses can initiate an array of unexpected physical responses, e.g. optoacoustics, ablation, \&c. In addition to bioeffects, optoacoustic spectroscopy is a very powerful detection method and considerable interest exist to develop a novel method for measuring Specific Absorption Rate (SAR) spanning the radio frequency band, specifically, ELF to terahertz.

\textbf{Aknowledgements.}\begin{small}\textit{ I am a military service member (or employee of the U.S. Government). This work was prepared as part of my official duties. Title 17 U.S.C. \S 105 provides that ’Copyright protection under this title is not available for any work of the United States Government.’ Title 17 U.S.C. \S 101 defines a U.S. Government work as a work prepared by a military service member or employee of the U.S. Government as part of that person’s official duties.}

\textit{The views, opinions and/or finding contained in this report are those of the authors and should not be construed as an official Department of the Navy, Air Force and Defense position, policy or decision unless so designated by other documentation. Trade names of materials and/or products of commercial or nongovernmental organizations are cited as needed for precision. These citations do not constitute official endorsement or approval of the use of such commercial materials and/or products.}

\textit{This work was funded by NAMRU-SA work unit number 2M8503.}
\end{small}


\bibliography{references}   
\bibliographystyle{spiejour}   

\end{document}